\documentclass[12pt]{article}
\usepackage{graphics}
\usepackage{bm}
\usepackage{graphicx}
\usepackage{amsmath}
\usepackage{amssymb}
\usepackage{amstext}
\usepackage{amscd}
\usepackage{amsfonts}
\usepackage{appendix}
\usepackage{hyperref}
\usepackage{textcomp}
\usepackage{epstopdf}
\usepackage{color}
\DeclareMathAlphabet{\EuFrak}{U}{euf}{m}{n}
\DeclareMathAlphabet{\EuScript}{U}{eus}{m}{n}

\newcommand{\nd}{\noindent}

\newcommand{\be}{\begin{equation}}
\newcommand{\ee}{\end{equation}}
\newcommand{\ben}{\begin{eqnarray}}
\newcommand{\een}{\end{eqnarray}}

\title{{\bf Quantum   treatment of  Verlinde's entropic force conjecture}}
\author{{\small{A. Plastino$^{1,4,5}$, M. C. Rocca$^{1,2,4}$,
G. L. Ferri $^3$}}, \\
\small{$^1$ Departamento de F\'{\i}sica,
Universidad Nacional de La Plata,}\\
\small{$^2$ Departamento de Matem\'{a}tica,
Universidad Nacional de La Plata,}\\
\small{$^3$Fac. de C. Exactas-National University La Pampa,} \\
\small{Peru y Uruguay, Santa Rosa, La Pampa, Argentina}\\
\small{$^4$ Consejo Nacional de Investigaciones Cient\'{\i}ficas
y Tecnol\'{o}gicas}\\
\small{(IFLP-CCT-CONICET)-C. C. 727, 1900 La Plata -
Argentina}\\\small{$^5$  SThAR - EPFL, Lausanne, Switzerland}}

\date{\today}

\begin{document}

\maketitle

\begin{abstract}
\nd Verlinde conjectured that gravitation is an emergent entropic force.
This surprising conjecture was proved in [Physica A {\bf 505} (2018) 190] within a purely
classical context. Here, we appeal to a quantum environment to deal with
the conjecture in the case of bosons and consider also the classical limit of quantum mechanics (QM).

\vskip 3mm \nd {\bf Keywords}: Gravitation, bosons, entropic force, emergent force, Verlide's conjecture..\\

\end{abstract}

\newpage

\tableofcontents

\newpage

\renewcommand{\theequation}{\arabic{section}.\arabic{equation}}

\section{Introduction}

\setcounter{equation}{0}

Eight years ago, Verlinde \cite{verlinde} proposed to link gravity with an entropic force. The ensuing conjecture was proved recently \cite{p1}, in a purely classical environment.\vskip 3mm

\nd According to Verlinde, gravity would emerge as a result of information about the positions of material particles,
connecting a thermal treatment of gravity to 't Hooft's holographic principle. In this perspective, gravitation should be regarded as  an emergent phenomenon. This Verlinde's idea was the focus of much attention. For
an example, see \cite{times,libro}. An excellent overview on the statistical mechanics of gravitation can be found in  Padmanabhan's article \cite{india}, and references therein.\vskip 2mm

\nd Verlinde's work originated endeavors in cosmology, the dark energy hypothesis, cosmological acceleration, cosmological inflation, and loop quantum gravity. The associated literature is extensive \cite{libro}.  An important contribution is that of Guseo \cite{guseo}. He showed that the local entropy function, related to a
logistic distribution, is a catenary and vice versa, an invariance that  may
be interpreted through  Verlinde's conjecture regarding gravity's  origin
 as an entropic force. Guseo advances a novel  interpretation of the local entropy in a system  \cite{guseo}.
 \vskip 2mm

\nd This paper does not treat any of these issues, though. Based on the fact that we proved Verlinde's hypothesis in a classical environment \cite{p1}, we wish to deal here with the quantal bosonic scenario.

\section{Entropic force for bosons}

\subsection{Quantum entropic force}

\setcounter{equation}{0}

The Bose gas' entropy is (see, for instance, \cite{lemons})
\begin{equation}
\label{eq2.1}
{\cal S}=Nk_B\left[\left(\frac {n} {N}\right)\ln\left(1+\frac {N} {n}\right)+
\ln\left(1+\frac {n} {N}\right)\right],
\end{equation}
where
\begin{equation}
\label{eq2.2}
n=V\left(\frac {E} {N}\right)^{\frac {3} {2}}
\left(\frac {4\pi e m} {2h^2}\right)^{\frac {3} {2}}.
\end{equation}
Since we can cast the volume as
\begin{equation}
\label{eq2.3}
V=\frac {4} {3}\pi r^3,
\end{equation}
we recast  (\ref{eq2.1})   as

\[{\cal S}=k_BV\left(\frac {E} {N}\right)^{\frac {3} {2}}
\left(\frac {4\pi em} {2h^2}\right)^{\frac {3} {2}}
\ln\left[1+\frac {N} {V}\left(\frac {N} {E}\right)^{\frac {3} {2}}
\left(\frac {2h^2} {4\pi em}\right)^{\frac {3} {2}}\right]+\]
\begin{equation}
\label{eq2.4}
Nk_B\ln\left[1+\frac {V} {N}\left(\frac {E} {N}\right)^{\frac {3} {2}}
\left(\frac {4\pi em} {2h^2}\right)^{\frac {3} {2}}\right].
\end{equation}
Now, according to \cite{p1} the entropic force is
\[F_e=-\lambda\frac {\partial S} {\partial A}=\]
\[\frac {\lambda 3k_BN} {8\pi r^2}
\frac {1} {1+\frac {3N} {4\pi r^3}\left(\frac {N} {E}\right)^{\frac {3} {2}}
\left(\frac {3h^2} {4\pi e m}\right)^{\frac {3} {2}}}-\]
\[\frac {\lambda k_B} {2}\left(\frac {E} {N}\right)^{\frac {3} {2}}
\left(\frac {4\pi e m} {2h^2}\right)^{\frac {3} {2}} r
\ln\left[1+\frac {3n} {4\pi r^3}\left(\frac {N} {E}\right)^{\frac {3} {2}}
\left(\frac {3h^2} {4\pi e m}\right)^{\frac {3} {2}}\right]-\]
\begin{equation}
\label{eq2.5}
\frac {\frac {\lambda k_B} {2}\left(\frac {E} {N}\right)^{\frac {3} {2}}
\left(\frac {4\pi e m} {2h^2}\right)^{\frac {3} {2}} r}
{1+\frac {4\pi r^3} {3N}\left(\frac {E} {N}\right)^{\frac {3} {2}}
\left(\frac {4\pi e m} {2h^2}\right)^{\frac {3} {2}}},
\end{equation}
where $\lambda$ is an arbitrary constant. We can recast the above expression  as
\[F_e=\frac {12\lambda k_BN\left(\pi emE\right)^{\frac {3} {2}} r}
{32\pi r^3\left(\pi emE\right)^{\frac {3} {2}}+
3^{\frac {5} {2}}N^{\frac {5} {2}}h^3}-\]
\[\frac {4\pi\lambda k_B\left(\pi emE\right)^{\frac {3} {2}}}
{\left(3N\right)^{\frac {3} {2}}h^3}r
\left\{\ln\left[32\pi r^3\left(\pi emE\right)^{\frac {3} {2}}+
(3N)^{\frac {5} {2}}h^3\right]-
\ln\left[32\pi r^3\left(\pi emE\right)^{\frac {3} {2}}\right]\right\}-\]
\begin{equation}
\label{eq2.6}
\frac {12\lambda k_BN\left(\pi emE\right)^{\frac {3} {2}} r}
{32\pi r^3\left(\pi emE\right)^{\frac {3} {2}}+
(3N)^{\frac {5} {2}}h^3},
\end{equation}
and then
\begin{equation}
\label{eq2.7}
F_e=\frac {4\pi\lambda k_B\left(\pi emE\right)^{\frac {3} {2}}}
{\left(3N\right)^{\frac {3} {2}}h^3}r
\left\{\ln\left[32\pi r^3\left(\pi emE\right)^{\frac {3} {2}}+
(3N)^{\frac {5} {2}}h^3\right]-
\ln\left[32\pi r^3\left(\pi emE\right)^{\frac {3} {2}}\right]\right\}.
\end{equation}

\subsection{Bosonic entropic force in the classical limit (CL)}

The CL is attained  for \cite{lemons}
\begin{equation}
\label{eq2.8}
\frac {N} {n}<<1,
\end{equation}
and in this limit the entropy becomes \cite{lemons}
\begin{equation}
\label{eq2.9}
{\cal S}=Nk_B\left[1+\ln\left(\frac {n} {N}\right)\right],
\end{equation}
or
\begin{equation}
\label{eq2.10}
{\cal S}=
\frac {5Nk_B} {2}+Nk_B\ln\left[\frac {V} {N}\left(\frac {E} {N}\right)^{\frac {3} {2}}
\left(\frac {4\pi m} {2h^2}\right)^{\frac {3} {2}}\right].
\end{equation}
Now, we have an entropic force of the form
\begin{equation}
\label{eq2.11}
F_e=-\lambda\frac {\partial S} {\partial A}=
-\frac {\lambda 3Nk_B} {8\pi r^2},
\end{equation}
which is indeed of the Newton appearance, so that Verlinde's conjecture gets proved in the classical limit.
 Note also that the entropic force (\ref{eq2.11}) can be derived as well from (\ref{eq2.7}) by taking $r$ large enough.

\begin{figure}[h]
\begin{center}
\includegraphics[scale=0.5,angle=0]{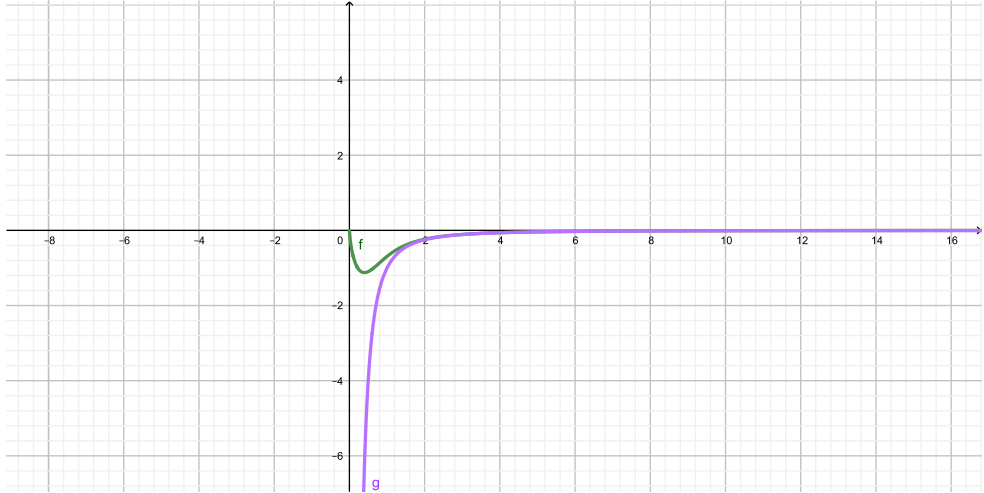}
\vspace{-0.2cm} \caption{Here we plot $F_e/C$, $C=\frac {3Nk_B\lambda} {8\pi}$.
Green line: Boson  entropic force.
Violet line: approximate semi-classic Bose-one.
Here, on the x axis, 1 unit=$10^{- 37}$ meter and, on the 
y axis, 1 unit=$10^{-4}$ Newton}\label{fig1}
\end{center}
\end{figure}
\nd In Fig. 1 we set:  $m$= the oxygen-molecule's mass,
$E=\frac {Nmv^2} {2}$, $v=1 meter/second$, and  $N$ is extracted from  
\[32\pi(\pi e mE)^{\frac {3} {2}}=(3N)^{\frac {5} {2}}h^3\], so that  
 $N=2.358458\times 10^{26}$. The classic entropic force  (\ref{eq2.11}) can also be encountered
starting from (\ref{eq2.7}) and taking $r$ large enough. Appealing such $r$ values and using
the equality  
\begin{equation}
\label{eq2.12}
-\frac {\lambda 3Nk_B} {8\pi r^2}=
-\frac {GmM} {r^2},
\end{equation}
where  $G$ is the gravitational constant, we detect the fact that  $\lambda=\lambda (m,M,N)$. 

\subsection{Entropic Potential Energy}

The entropic force is proportional to the derivative of the entropy with respect to 
the spherical area $A$.  It is interesting  to calculate 
the corresponding potential energy $E_P$. To such an end we define the new constants $a$ and $b$ in the fashion 
$a=(3N)^{\frac {5} {2}}h^3$ and $b=32\pi(\pi e mE)^{\frac {3} {2}}$.
Using  reference \cite{gra} we can compute the potential energy we are looking for from the entropic force's expression. The ensuing calculation is simple but lengthy. One arrives at the result 
\[E_P(r)=\frac {3Nk_B\lambda} {8\pi}\frac {b} {a}\left\{\frac {r^2} {2}
\ln\left(1+\frac {a} {br^3}\right)-
\frac {a^{\frac {2} {3}}} {2b^{\frac {2} {3}}}\left\{\frac {1} {2}\ln
\left[\frac {\left[r+\left(\frac {a} {b}\right)^{\frac {1} {3}}\right]^2} 
{r^2-\left(\frac {a} {b}\right)^{\frac {1} {3}}r+\left(\frac {a} {b}\right)^{\frac {2} {3}}}
\right]\right.\right.+\]
\begin{equation}
\label{eq2.13}
\left.\left.\sqrt{3}\left[\frac {\pi} {2}-
\arctan\left[\frac{2r-\left(\frac {a} {b}\right)^{\frac {1} {3}}}
{\sqrt{3}\left(\frac {a} {b}\right)^{\frac {1} {3}}}\right]\right]
\right\}\right\},
\end{equation}
where we have set $E_P(r)=0$ for $r\rightarrow\infty$.
For $r$ large the potential energy adopts the appearance
\begin{equation}
\label{eq2.14}
E_P(r)=
-\frac {\lambda 3Nk_B} {8\pi r},
\end{equation}
which is consistent with the result  (\ref{eq2.11}).

\section{Conclusions}

We have here  considered Verlinde's [entropic force - Gravitation] link, proved recently in a classical context \cite{p1}, in a quantum bosonic scenario.  We have seen that Verlinde's conjecture holds in this scenario as well. Further, the quantum emergent gravitation \`a la Verlinde does not diverge at the origin. Such an  asymptotic behaviour is rather surprising. One wonders whether this emergent  gravitation-behaviour might perhaps be an artifact of not being able to include general relativity effects.  Moreover, in two limits

\begin{itemize}

\item the classical limit of QM  
\item   $r \rightarrow \infty$,

\end{itemize}
the Newton $r$-dependence of the gravitation force is recovered. A natural challenge is to attack the fermionic case. This we did in reference \cite{fer}. For fermions, the vanishing of the entropic force at the origin is also observed.

\setcounter{equation}{0}

\end{document}